\documentstyle[12pt,epsf]{article}
\def\chapter{\clearpage         % Starts new page.
   \global\@topnum\z@           % Prevents figures from going at top of page.
   \@afterindenttrue            % Suppresses indent in first paragraph.  Change
   \secdef\@chapter\@schapter}  % to \@afterindenttrue to have indent.
\begin{document}
\begin{titlepage}
%\begin{flushright}
%KEK-preprint-94-*\\
%NWU-HEP 94-07\\
%DPNU-94-59\\
%TIT-HPE-94-013\\
%TUAT-HEP 94-07\\
%OCU-HEP 94-07\\
%PU-94-692\\
%INS-REP 1077\\
%KOBE-HEP 94-06\\
%TU-HEP 94-*\\
%\end{flushright}

%\special{psfile=kekmark.ps hscale=0.7 vscale=0.7 hoffset=-150 voffset=-220}

\begin{picture}(350,130)(0,10)
  \put(330,135){KEK-preprint-94-162}
  \put(330,120){NWU-HEP 94-07}
  \put(330,105){DPNU-94-59}
  \put(330,90){TIT-HPE-94-013}
  \put(330,75){TUAT-HEP 94-07}
  \put(330,60){OCU-HEP 94-07}
  \put(330,45){PU-94-692}
  \put(330,30){INS-REP 1077}
  \put(330,15){KOBE-HEP 94-06}
  \put(-10,30){\epsfysize=3.5cm\epsfbox{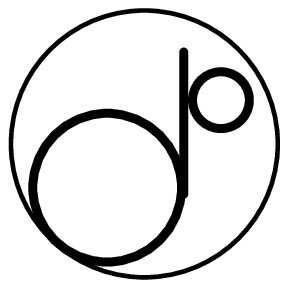}}
\end{picture}

\begin{center}
\begin{Large}
%Test of MLLA calculation of particle spectra in the $e^+e^-$
%  annihilation at $\sqrt{s}$=58GeV
Measurement of inclusive particle spectra and test of MLLA prediction
in $e^+e^-$ annihilation at $\sqrt{s}$=58GeV
\end{Large}

\vskip 0.5cm
(TOPAZ collaboration)
\vskip 0.3cm

R.Itoh$^a$, M.Yamauchi$^a$, A.Yamaguchi$^b$
K.Abe$^c$, T.Abe$^c$, I.Adachi$^a$,
K.Adachi$^b$, M.Aoki$^c$, M.Aoki$^d$, S.Awa$^b$,
K.Emi$^e$, R.Enomoto$^a$, H.Fujii$^a$, K.Fujii$^a$,T.Fujii$^f$,
J.Fujimoto$^a$,
 K.Fujita$^g$, N.Fujiwara$^b$, H.Hayashii$^b$,
B.Howell$^h$, N.Iida$^b$, Y.Inoue$^g$, H.Iwasaki$^a$, M.Iwasaki$^b$,
K.Kaneyuki$^d$, R.Kajikawa$^c$,
S.Kato$^i$, S.Kawabata$^a$, H.Kichimi$^a$, M.Kobayashi$^a$,
 D.Koltick$^h$, I.Levine$^h$, S.Minami$^d$,
K.Miyabayashi$^c$, A.Miyamoto$^a$, K.Muramatsu$^b$, K.Nagai$^j$,
K.Nakabayashi$^c$, E.Nakano$^c$, O.Nitoh$^e$, S.Noguchi$^b$, A.Ochi$^d$,
F.Ochiai$^k$,
 N.Ohishi$^c$, Y.Ohnishi$^c$, Y.Ohshima$^d$,
H.Okuno$^i$, T.Okusawa$^g$,T.Shinohara$^e$, A.Sugiyama$^c$,
S.Suzuki$^c$, S.Suzuki$^d$, K.Takahashi$^e$, T.Takahashi$^g$,
 T.Tanimori$^d$, T.Tauchi$^a$, Y.Teramoto$^g$, N.Toomi$^b$,
 T.Tsukamoto$^a$, O.Tsumura$^e$, S.Uno$^a$, T.Watanabe$^d$,
Y.Watanabe$^d$ and A.Yamamoto$^a$ \\

\end{center}
{\small \it
\leftline{(a)
    KEK, National Laboratory for High Energy Physics, Ibaraki-ken 305,
    Japan }
\leftline{(b)
    Department of Physics, Nara Women's University, Nara 630, Japan }
\leftline{(c)
    Department of Physics, Nagoya University, Nagoya 464, Japan}
\leftline{(d)
    Department of Physics, Tokyo Institute of Technology, Tokyo 152,
    Japan}
\leftline{(e)
    Department of Applied Physics, Tokyo Univ. of Agriculture and
    Technology, Tokyo 184, Japan}
\leftline{(f)
    Department of Physics, University of Tokyo, Tokyo 113, Japan}
\leftline{(g)
    Department of Physics, Osaka City University, Osaka 558, Japan }
\leftline{(h)
    Department of Physics, Purdue University, West Lafayette, IN
    47907, USA }
\leftline{(i)
    Institute for Nuclear Study, University of Tokyo, Tanashi,
     Tokyo 188, Japan }
\leftline{(j)
    The Graduate School of Science and Technology, Kobe University,
    Kobe 657,
    Japan }
\leftline{(k)
    Faculty of Liberal Arts, Tezukayama University, Nara 631, Japan }
}
\begin{center}
{\it Submitted to Physics Letters B}
\end{center}
\end{titlepage}

\newpage
\begin{titlepage}
\begin{abstract}
Inclusive momentum spectra are measured for all charged particles and for
each of $\pi^{\pm}$, $K^{\pm}$, $K^0/\overline{K^0}$, and
$p/\overline{p}$
in hadronic events produced via $e^+e^-$ annihilation at
$\sqrt{s}$=58GeV . The measured
spectra are compared with QCD predictions based on the
modified leading log approximation(MLLA).
The MLLA model reproduces the measured spectra
well. The energy dependence of the peak positions of the
spectra is studied by
comparing the measurements with those at other energies.
The energy dependence is also well described by the MLLA model.
\end{abstract}
\end{titlepage}

\newpage
\section{Introduction}

       The LLA (leading log approximation) parton shower model
well reproduces the various
distributions of observables for hadronic final states of $e^+e^-$
collisions, when the ``coherence'' effect of soft gluons
is taken into account.
Several Monte Carlo programs were
written based on this scheme (JETSET63\cite{JETSET63}, for example) \
and were used in various experiments to
study QCD. However, since the coherence effect is the consequence
of higher order corrections, the
effect could only be inserted by hand in the LLA scheme in these Monte
Carlo programs.

On the other hand, even in low $Q^2$ region,
the momentum spectrum of gluons in the parton shower process can be
analytically calculated
using the modified leading log approximation(MLLA)\cite{MLLA}. The
coherence effect is taken into account in MLLA by consistently
importing a part of next-to-leading order corrections.
The distribution of the particle spectra is expressed as a function of
two parameters,
$Y$=log($\frac{E\Theta}{Q_0}$) and $\lambda$=log($\frac{Q_0}{\Lambda}$):

\begin{eqnarray}
x_p{\overline D^g_q}(x,Y,\lambda) &=& \frac{4C_F(Y+\lambda)}{bB(B+1)}
\int_{\epsilon +{\rm i}\infty}^{\epsilon -{\rm
i}\infty}\frac{d\omega}{2\pi{\rm i}}x_p^{-\omega}\Phi
(-A+B+1,B+2,-\omega (Y+\lambda)) \nonumber\\
&\times& \frac{\Gamma (A)}{\Gamma (B)}
(\omega\lambda)^B\Psi (A,B+1,\omega\lambda) \label{eq:MLLA}
\end{eqnarray}
Here $x_p$ is the momentum of a particle normalized by the beam energy
$E=\sqrt{s}/2$, $\Theta$ is the opening angle of the jet cone,
and $C_F$=$\frac{4}{3}$,  $b=\frac{23}{3}$, $A$=$\frac{12}{b\omega}$ and
$B=\frac{307}{27b}$ for five quark flavors, respectively.
The two functions, $\Phi$ and $\Psi$, are two solutions of the
confluent hyper-geometric equation.
The quantities $\Lambda$ and $Q_0$ are the QCD scale parameter and the
energy cut-off of the parton evolution, respectively.

This calculation predicts depletion of soft partons as a consequence of
the destructive interference of soft gluons.
This depletion shows up clearly in eq.~\ref{eq:MLLA} when written as a
function of $\xi = {\rm ln}(1/x_p)$.
This function has a maximum at a certain $\xi$ value
and decreases in the larger $\xi$ region. The coherence effect reduces the
available phase space in this region and therefore the effect can be
studied by comparing the measured inclusive cross section with this
calculation.

However, since this expression is calculated for partons, it cannot be
compared directly with the measurement unless the distribution at the
level of
final state hadrons is ensured to be similar to that of partons.
The concept of Local Parton Hadron Duality (LPHD) dictates that the
distribution of the final state hadrons is closely related to that of
partons\cite{LPHD}.
The conversion of partons into hadrons,
which occurs at low virtuality scale, includes only small momentum
transfer, and hence leaves the distribution essentially unchanged.
In this article we compare the calculated parton spectrum (eq.~\ref{eq:MLLA})
directly with the measured hadron spectrum assuming LPHD.

The $Q_0$, which was primarily introduced to regularize collinear singularity,
gives a cut-off on parton energies.
Therefore, the value of $Q_0$ should be close to the mass of
the particle being considered\cite{mh}, while $\Lambda$ should be
common to all the particle species.
These dependences can be experimentally tested by measuring the values
of $Q_0$ and $\Lambda$ for various particle species.

The momentum spectrum might still be distorted by fragmentations and
decays in spite of LPHD. This complication
can be avoided partially when the energy evolution of the distribution
is considered.
The peak position of the distribution (eq.~\ref{eq:MLLA}) is given
using the limiting spectra calculation in which $\Lambda$ is
assumed to be equal to $Q_0$\cite{MLLA}:
\begin{equation}
\xi_{max} = \frac{1}{2}Y
+B\sqrt{\frac{b}{16N_c}Y} - \frac{bB^2}{16N_c}
\label{eq:Peak}
\end{equation}
where $N_c$ is the number of colors ($= 3$).
This should be compared with the
variation as $\xi_{max}\sim Y$
expected from phase space consideration alone.

%whereas it varies as $\xi_{max}\sim \frac{1}{2}\log E$ in this expression.
%Therefore the measurement of the logarithmic slope gives a test of the
%prediction.

\section{Event Selection}

  % sqrt_s and the integrated luminosity
The data used in this analysis are accumulated by the TOPAZ detector
at the TRISTAN
$e^+e^-$ collider at center-of-mass energies between 52.0 and 61.4~GeV.
The average energy is 58.0~GeV, and the total integrated luminosity is
113.7/pb.

  % relevant detectors and trigger condition
Details of the TOPAZ detector is described elsewhere\cite{TOPAZ}.
In this analysis,
the data from a Time Projection Chamber (TPC) is mainly used
for tracking and dE/dx measurement for charged particles\cite{TPC}.
The trigger condition relevant to this analysis is the track trigger which
requires two or more charged tracks in the fiducial volume of the TPC, and the
energy trigger which requires 4~GeV or more energy deposit in the barrel lead
glass calorimeter.
The event trigger is generated by a logical OR of those two, and the trigger
efficiency for multihadronic events is practically 100\%.

  % hadron event selection
Out of the triggered events, multihadronic events are selected
by the following
conditions; (a) five or more charged tracks having transverse momenta with
respect to the beam axis larger than 0.15~GeV/$c$ are
originating from the
interaction point with the angle larger than 37$^\circ$ with respect
to the beam axis,
(b) total visible energy is larger than 1/2 of the
center-of-mass energy, and (c) momentum imbalance along the beam direction is
smaller than 0.4.
By these conditions, the event selection efficiency is estimated to be
67.1\% with a background contamination of less than 2.0\%.
Two-photon processes and $\tau^+\tau^-$ productions are the
main sources of the background.
In addition to these cuts, the jet axis is required
to have an angle larger than 40$^\circ$ with
respect to the beam axis to ensure that the event is
well contained in the detector acceptance.
Applying these criteria, 11247 events remain to be
used in the analysis.

\section{Particle Identification and Measurement of Cross section}

\begin{figure}
{\centerline{\epsfysize=15cm\epsfbox{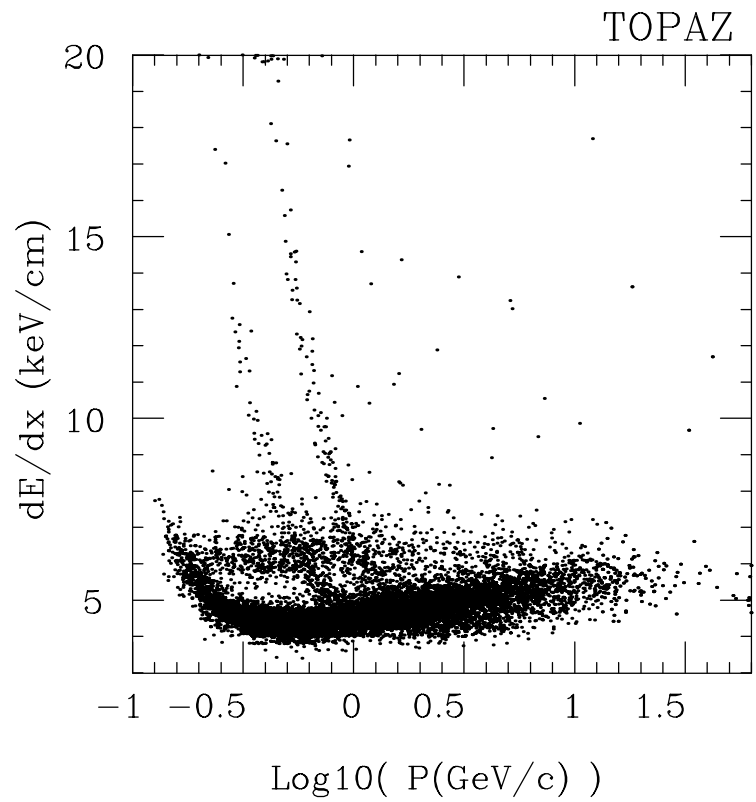}}}
\caption{dE/dx distribution as a function of track momentum measured
  by TOPAZ-TPC.\label{fig:dEdX}}
\end{figure}

The particle species $\pi^{\pm}$, $K^{\pm}$ and
$p/\overline{p}$ are identified
by measuring the dE/dx of tracks detected in TPC. The detail of the
particle identification technique is described in ref.\cite{Itoh-D}.
The typical resolution of the dE/dx measurement for minimum ionizing
pions is 4.6\%. Fig.~\ref{fig:dEdX} shows dE/dx distribution
as a function of the track momentum for a part of the
event sample.

\begin{figure}[t]
{\centerline{\epsfysize=10cm\epsfbox{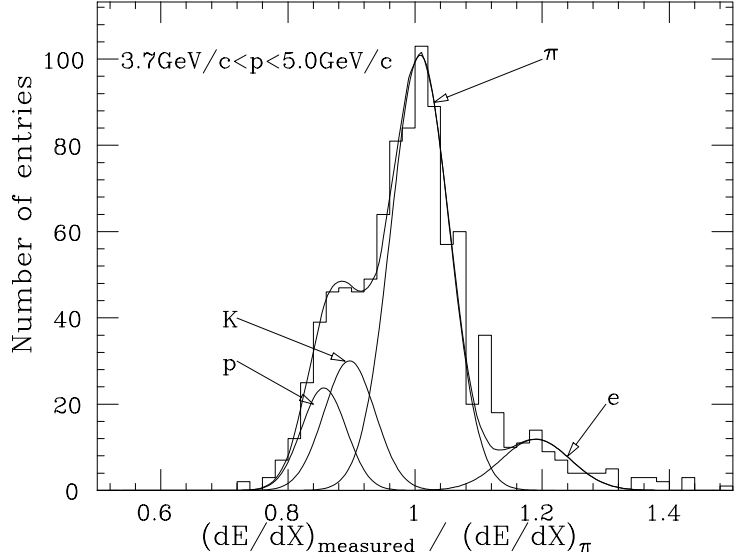}}}
\caption{A typical fit to dE/dx distribution in a momentum bin. The
  distribution is fitted by four Gaussians corresponding to $e^{\pm}$,
  $\pi^{\pm}$, $K^{\pm}$ and $p/\overline{p}$. \label{fig:dEdXfit}}
\end{figure}

The cross section is calculated from the
number of tracks in each momentum slice.
In the low momentum region, the number of tracks for each particle species are
directly obtained by counting the number of tracks in each dE/dx band.
The dE/dx bands are, however, not well separated in the higher momentum
region. To extract the number of
each particle species in this region,
the dE/dx distribution in a momentum slice is fitted
with a superposition of four Gaussians
corresponding to $e^\pm$, $\pi^\pm$'s, $K^\pm$'s and
$p/\overline{p}$ in each momentum slice.
The widths and the centers of the Gaussians are determined from the
measurement for Bhabha events and cosmic ray $\mu$'s.
Only the normalizations of the Gaussians are the free parameters of
the fit. A typical fit to the dE/dx distribution is shown in
Fig.~\ref{fig:dEdXfit}.
Because of a hardware calibration problem of the TPC, the dE/dx
resolution is time dependent. Therefore the
data sample is divided into two groups and the procedure to count the
number of tracks of each particle species is done separately for these two
groups.

  % Kshort

$K_s$'s are identified by searching for their daughter charged pion pairs
in TPC. A pair of tracks must satisfy following conditions to be
identified as a $K_s$: a) the distance of the closest
approach of the two tracks is less than 0.8cm; b) the distance from the
interaction point to the decay vertex is longer than 2.0cm where the decay
vertex is defined as the center of the closest approach of two tracks;
c) the angle formed by the vector from the interaction point to the
decay vertex and the momentum sum of
the two tracks at
the decay vertex is less than 8 degrees if the position of decay vertex
is longer than 6.0 cm; or c') the closest distance from the vector sum
to the interaction point is less than 0.6 cm if the decay length is
shorter than 6.0 cm; d) the  closest approach of one of the tracks to the
interaction point is more than 0.5cm if the momentum sum of
the tracks is less than 2GeV/c; e) tracks are identified as
$\pi^\pm$ by the dE/dx measurement in the TPC; and f) the tracks are not
from the gamma conversion.

\begin{figure}[t]
{\centerline{\epsfysize=12cm\epsfbox{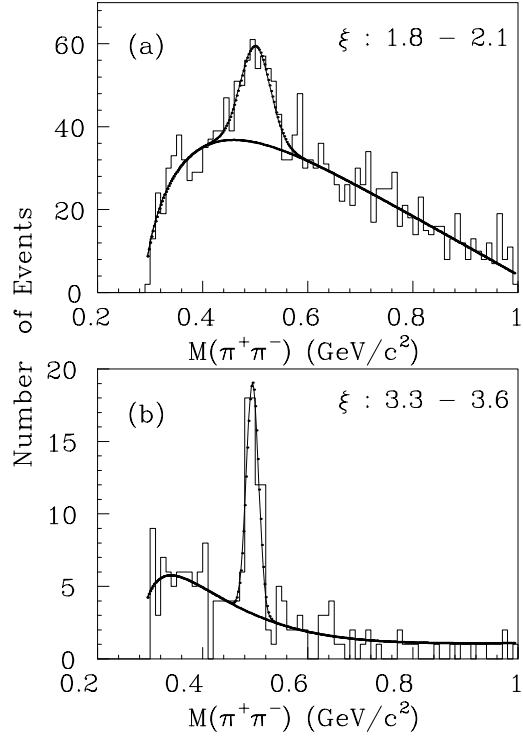}}}
\vspace{0.5cm}
\caption{Reconstructed mass distributions of $K_s$ in two different
  momentum bins: (a) 1.8$<\xi<$2.1, (b) 3.3$<\xi<$3.6 \label{fig:Ks-mass}}
\end{figure}

For each of track pairs remained after these selections, the invariant mass is
calculated assuming the tracks are charged pions.
Fig.~\ref{fig:Ks-mass} shows the mass distribution for the reconstructed
$K_s$'s. The distribution is fitted by a sum of a Gaussian and a background
function (exponential + polynomial) and the number of reconstructed $K_s$'s
is obtained to be 771 $\pm$ 61. In the same way, the numbers of these
$K_s$'s are counted for each momentum
slice.

  % acceptance correction
The inclusive cross section, $1/\sigma_{had}\;d\sigma/dx_p$ is then
calculated from the number of particles observed in each momentum slice
using following formula:
\begin{equation}
\frac{1}{\sigma_{had}} \frac{d\sigma}{dx_i} =
A(x_i)\frac{1}{\Delta x_i} \frac{N_{obs}(x_i)}{N_{had}},
\end{equation}
where $x_i$ is the i'th slice of the momentum fraction,
$\sigma_{had}$ is the total hadronic cross section,
$\Delta x_i$ is the width of the slice,
$N_{had}$ is the number of hadronic events in the event sample, and
$N_{obs}(x_i)$ is the number of particles counted in the slice, respectively.
$A(x_i)$ is the factor to correct the counted number for the effects of
detector acceptance and initial state radiation. $A(x_i)$ is
calculated for each momentum slice separately with the Monte Carlo
simulation as
\begin{equation}
A(x_i) = \frac{N_{gen}(x_i)}{N_{gen}^{total}} /
\frac{N_{sim}(x_i)}{N_{sim}^{total}}
\end{equation}
where $N_{gen}(x_i)$ is the generated number of particles in the
slice $x_i$
without including the effects of the detector
acceptance and the initial state radiation. $N_{gen}^{total}$ is the
corresponding
total number of generated events. $N_{sim}(x_i)$ is the number of detected
particles in the slice $x_i$ generated with the effects
of detector acceptance and initial state radiation, and
$N_{sim}^{total}$ is the corresponding total number of generated events.
The JETSET6.3\cite{JETSET63} and 7.3\cite{JETSET73} Monte Carlo programs are
used to obtain these numbers combined with the
TOPAZ detector simulation program. As for $K_s$, to convert the counted
number to the cross section of $K^0/\overline{K^0}$, $A(x_i)$ is
multiplied by 2.

The cross sections are measured as a function of $\xi$.
%where $\xi$ is
%defined as $\xi = {\rm ln} (1/x_p)$. $x_p$ is the momentum fraction of a
%particle.
The measured cross section for all charged particles is
shown in Table~\ref{Table:Xsec0}. Table~\ref{Table:Xsec1} shows the
cross sections measured for each of
$\pi^{\pm}$, $K^{\pm}$, $p/\overline{p}$,
while that for $K^0/\overline{K^0}$ is shown in
Table ~\ref{Table:Xsec2}.

\begin{table}
\begin{center}
\begin{tabular}{||c|c||}
\hline
$\xi$ = ln(1/$x_p$) & $1/\sigma_{tot}\;d\sigma/d\xi$ \\ \hline
  0.7  &    0.287$\pm$0.015 \\
  0.9  &    0.656$\pm$0.018 \\
  1.1  &    0.942$\pm$0.024 \\
  1.3  &    1.313$\pm$0.031 \\
  1.5  &    1.783$\pm$0.041 \\
  1.7  &    2.455$\pm$0.055 \\
  1.9  &    3.116$\pm$0.067 \\
  2.1  &    3.547$\pm$0.076 \\
  2.3  &    4.182$\pm$0.089 \\
  2.5  &    4.380$\pm$0.093 \\
  2.7  &    4.877$\pm$0.103 \\
  2.9  &    5.089$\pm$0.107 \\
  3.1  &    5.674$\pm$0.119 \\
  3.3  &    5.705$\pm$0.120 \\
  3.5  &    5.673$\pm$0.119 \\
  3.7  &    5.406$\pm$0.114 \\
  3.9  &    5.135$\pm$0.109 \\
  4.1  &    4.564$\pm$0.098 \\
  4.3  &    4.037$\pm$0.087 \\
  4.5  &    3.265$\pm$0.071 \\
  4.7  &    2.498$\pm$0.055 \\
  4.9  &    1.637$\pm$0.046 \\
\hline
\end{tabular}
\caption{Measured cross section for all charged particles as a
  function of $\xi$. Errors include both of statistical and systematic
  errors. \label{Table:Xsec0}}
\end{center}
\end{table}

\begin{table}
\begin{center}
\begin{tabular}{||c|c|c|c||}
\hline
 & \multicolumn{3}{|c|}
{$1/\sigma_{tot}\;d\sigma/d\xi$} \\ \hline
$\xi$ = ln(1/$x_p$) & $\pi^{\pm}$ & $K^{\pm}$ & $p/\overline{p}$ \\
\hline
   1.32  &     1.07$\pm$0.12 & 0.41$\pm$0.08 & 0.17$\pm$0.04 \\
   1.62  &     1.50$\pm$0.14 & 0.61$\pm$0.10 & 0.25$\pm$0.05 \\
   1.92  &     2.24$\pm$0.21 & 0.63$\pm$0.10 & 0.23$\pm$0.04 \\
   2.22  &     2.75$\pm$0.23 &       -       &       -       \\
   2.47  &     3.25$\pm$0.27 &       -       &       -       \\
   2.62  &     3.75$\pm$0.42 & 0.75$\pm$0.19 & 0.30$\pm$0.21 \\
   2.77  &     3.95$\pm$0.64 & 0.69$\pm$0.16 & 0.33$\pm$0.22 \\
   2.98  &     3.84$\pm$0.80 &       -       &       -       \\
   3.22  &         -         &       -       &       -       \\
   3.41  &     4.75$\pm$0.50 & 0.51$\pm$0.08 & 0.22$\pm$0.04 \\
   3.51  &     4.86$\pm$0.46 & 0.39$\pm$0.06 & 0.20$\pm$0.03 \\
   3.61  &     4.79$\pm$0.37 & 0.44$\pm$0.06 & 0.21$\pm$0.03 \\
   3.77  &     4.53$\pm$0.37 & 0.48$\pm$0.06 & 0.13$\pm$0.02 \\
   3.96  &     4.18$\pm$0.34 & 0.29$\pm$0.04 &       -       \\
   4.14  &     3.82$\pm$0.36 & 0.24$\pm$0.03 &       -       \\
   4.42  &     3.27$\pm$0.28 & 0.12$\pm$0.02 &       -       \\
   4.71  &     2.27$\pm$0.18 &       -       &       -       \\
   4.83  &     1.94$\pm$0.17 &       -       &       -       \\
\hline
\end{tabular}
\caption{Measured cross sections for $\pi^{\pm}$, $K^{\pm}$ and
  $p/\overline{p}$ as functions of
  $\xi$. Special non-uniform binning of $\xi$ is used to optimize the particle
  identification by the dE/dx measurement.
  Errors include both of statistical and systematic errors.
\label{Table:Xsec1}}
\end{center}
\end{table}

\begin{table}
\begin{center}
\begin{tabular}{||c|c||}
\hline
$\xi$ = ln(1/$x_p$) & $1/\sigma_{tot}\;d\sigma(K_s)/d\xi$ \\ \hline
    1.80 &   0.58$\pm$0.10 \\
    2.25 &   0.65$\pm$0.12 \\
    2.55 &   0.68$\pm$0.10 \\
    2.85 &   0.61$\pm$0.09 \\
    3.15 &   0.62$\pm$0.09 \\
    3.45 &   0.49$\pm$0.09 \\
    3.80 &        -        \\
    4.25 &        -        \\
    4.05 &   0.23$\pm$0.04 \\
\hline
\end{tabular}
\caption{Measured cross section for $K^0/\overline{K^0}$ as a
  function of $\xi$. Errors are statistical only.\label{Table:Xsec2}}
\end{center}
\end{table}

The systematic error in the measurement is estimated by considering
following sources. One is the systematic ambiguity in the acceptance
correction. This is studied by varying the fragmentation parameters in
the Monte Carlo programs and by turning on and off the energy loss and
the nuclear interaction effects in the detector simulation program.
The ambiguity is estimated to be 2\% for all charged particles while 3\% for
$\pi^{\pm}$, $K^{\pm}$ and $p/\overline{p}$.
For $\pi^{\pm}$, $K^{\pm}$ and $p/\overline{p}$,
the ambiguity in the dE/dx curve as a function of
momentum used to extract numbers of each particle species
is also a source of the systematic error.
This is estimated
to be 5\% for $\pi^{\pm}$ and 10\% for $K^{\pm}$ and $p/\overline{p}$.

\section{Comparison with MLLA prediction}

\begin{figure}[t]
{\centerline{\epsfysize=10cm\epsfbox{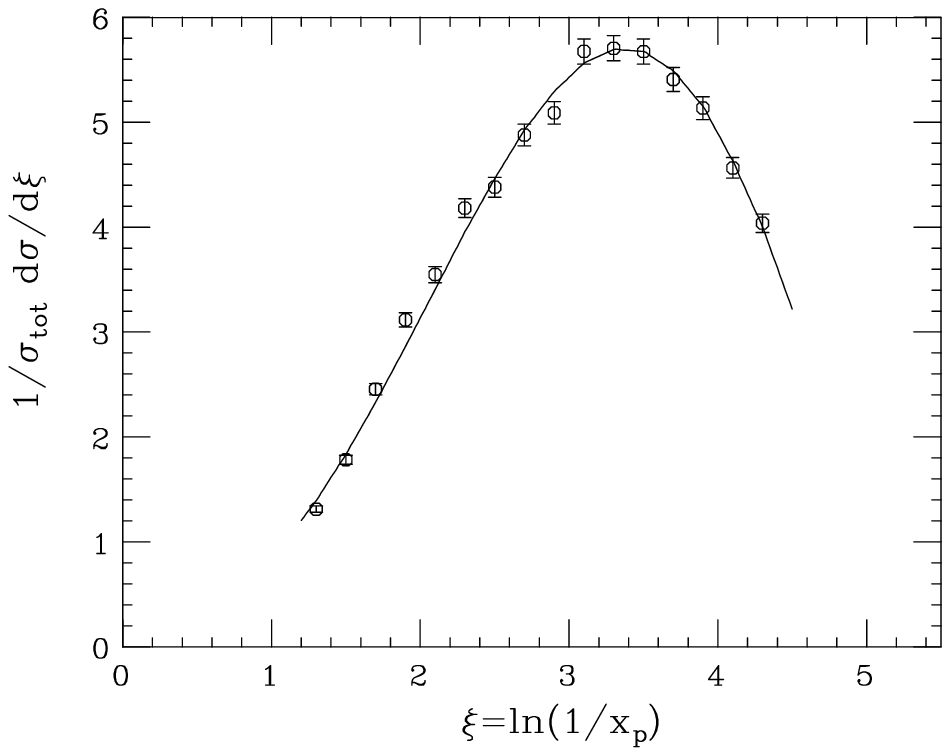}}}
\caption{The cross section measured for all charged particles as a function of
$\xi=ln(1/x_p)$. Solid curve shows the fitted MLLA
calculation.\label{Charged_Xsec}}
\end{figure}

        The measured particle spectra are then compared directly with
the MLLA predictions by assuming LPHD.
Since the $\Lambda$ and $Q_0$ in the MLLA calculation
are unknown, their values are determined by fitting the MLLA
formula of the cross section to the data.
Fig.~\ref{Charged_Xsec} shows the measured cross section for all
charged particles as a function of $\xi$ together with the result of
the best fit
shown in the solid line. As seen from Fig.~\ref{Charged_Xsec},
the fitted MLLA calculation well
reproduces the measured cross section.

\begin{figure}
{\centerline{\epsfysize=15cm\epsfbox{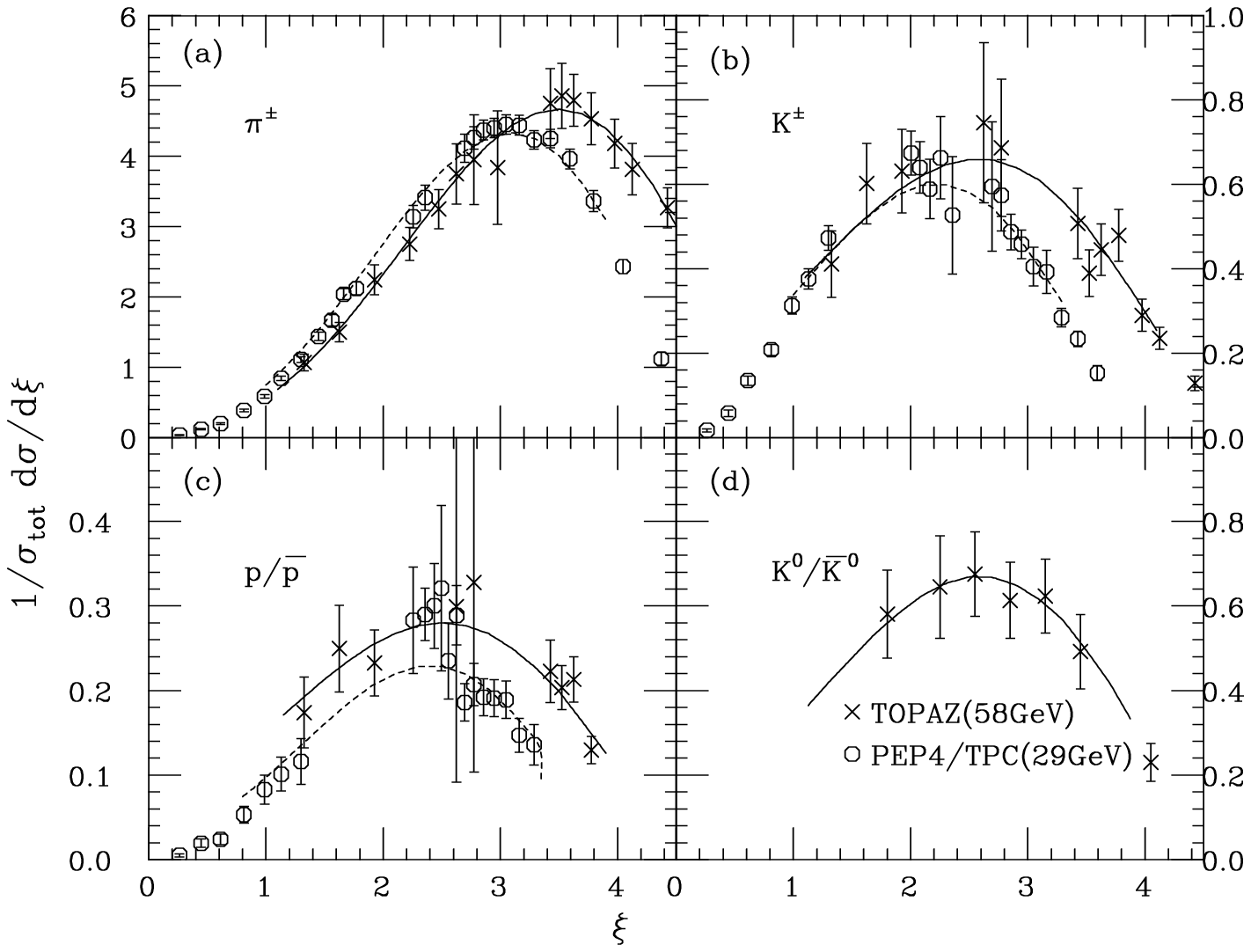}}}
\caption{The cross sections measured for (a) $\pi^{\pm}$, (b) $K^{\pm}$,
  (c) $p/\overline{p}$ and (d) $K^0/\overline{K^0}$.
  Both of TOPAZ and PEP4/TPC measurements
  are plotted with MLLA fits (Solid curve:TOPAZ, Dashed curve:PEP4).
  \label{pikp_Xsec}}
\end{figure}

Fig.~\ref{pikp_Xsec} shows the cross sections measured for
$\pi^{\pm}$, $K^{\pm}$, $K^0/\overline{K^0}$ and $p/\overline{p}$
with the
fitted MLLA calculations. Also shown are the measurements by
PEP4/TPC\cite{PEP4} at $\sqrt{s} = 29 {\rm GeV}$ for $\pi^{\pm}$,
$K^{\pm}$, and $p/\overline{p}$. These measurements
are also fitted by the MLLA calculations. The fit is performed in the
range where the numerical calculation of the MLLA function is
reliable. The range is typically 1.0$<\xi<$4.0.
The measured cross sections
are well reproduced by MLLA. The figure demonstrates the
energy evolution of the peak position of the spectra:
the peak position in $\xi$ becomes larger in TOPAZ
measurements at $\sqrt{s}=58{\rm GeV}$  than those of PEP4/TPC at
$\sqrt{s}=29 {\rm GeV}$ regardless of particle species. This is also
predicted by the MLLA calculation. The detail of the study of the evolution
of the peak position is described in the next section.

Tables~\ref{table-MLLA} and \ref{table-MLLA-2}
summarize the obtained values of $\Lambda$
and $Q_0$. Two different fits are performed: one is to determine both of
$\Lambda$ and $Q_0$ from the fit, and the other is to determine
$Q_0$ from the fit with $\Lambda$ fixed at 200 MeV.
In both cases, the value of $Q_0$ becomes larger for heavier particles.
Meanwhile, in the first case, the determined value of $\Lambda$
stays at relatively lower values (100-300 MeV). This result gives a
strong support to the MLLA + LPHD conjecture.

\begin{table}
\begin{center}
\begin{tabular}{||c|c|c|c|c||}
\hline
  & \multicolumn{2}{|c|}{TOPAZ(58GeV)} &
  \multicolumn{2}{|c|}{PEP4(29GeV)} \\ \hline
Free fit& $\Lambda$ (MeV) & $ Q_0$(MeV)
   & $\Lambda$ (MeV) & $ Q_0$(MeV) \\ \hline
all charged & 291$\pm$10 & 375$\pm$8 & - & - \\
$\pi^{\pm}$ & 281$\pm$20 & 339$\pm$25 & 270$\pm$7 & 250$\pm$7 \\
$K^{\pm}$ & 118$\pm$68 & 575$\pm$80 & 243$\pm$74 & 531$\pm$80 \\
$K^{0}/\overline{K^0}$ & 185$\pm$19 & 649$\pm$76 & - & - \\
$p/\overline{p}$ & 143$\pm$81 & 657$\pm$90 & 356$\pm$105 & 531$\pm$100 \\
\hline
\end{tabular}
\caption{$\Lambda$ and $Q_0$ for each of particle species determined
from the fits with both $\Lambda$ and $Q_0$ as free parameters.
\label{table-MLLA}}
\end{center}
\end{table}

\begin{table}
\begin{center}
\begin{tabular}{||c|c|c||}
\hline
  & TOPAZ(58GeV) & PEP4(29GeV) \\ \hline
$\Lambda=200$MeV & $Q_0$ (MeV) & $Q_0$ (MeV) \\ \hline
all charged & 297$\pm$12 & - \\
$\pi^{\pm}$ & 275$\pm$12 & 217$\pm$6 \\
$K^{\pm}$ & 588$\pm$44 & 514$\pm$45 \\
$K^{0}/\overline{K^0}$ & 663$\pm$110 & - \\
$p/\overline{p}$ & 627$\pm$92 & 390$\pm$ 50 \\ \hline
\end{tabular}
\caption{$Q_0$ for each of particle species determined by the fits with
$\Lambda$ fixed at 200MeV.
\label{table-MLLA-2}}
\end{center}
\end{table}

\section{The Energy Evolution of $\xi$ Distribution}

The energy evolution of the peak positions in $\xi$ is studied by comparing
the measurements at $\sqrt{s}$=29GeV(PEP4/TPC), 58GeV(TOPAZ)
and 91GeV(ALEPH)\cite{ALEPH} as a part of the PTA project\cite{PTA}.
Table~\ref{Peak_position} shows the result. The peak positions and
their errors are estimated using the fitted MLLA functions for PEP4
and TOPAZ measurements.
The MLLA predictions
using the limiting spectra calculation (eq.~\ref{eq:Peak})
with $\Lambda=200MeV$ are also shown.
As seen, the peak positions for light particles
(all charged particles and
$\pi^{\pm}$) agree with the limiting spectra calculation.
The peak positions become smaller for heavier
particles regardless of the energy scale.
% except for that for protons measured by ALEPH.

\begin{table}
\begin{center}
\begin{tabular}{||c|c|c|c||}
\hline
  & PEP4/TPC(29GeV) & TOPAZ(58GeV) & ALEPH(91GeV) \\ \hline
all charged & - & 3.30$\pm$0.05 & 3.61$\pm$0.01 \\
$\pi^{\pm}$ & 3.10$\pm$0.12 & 3.48$\pm$ 0.18 & 3.81$\pm$0.02 \\
$K^{\pm}$ & 2.30$\pm$0.20 & 2.56$\pm$ 0.29 & 2.63$\pm$0.04 \\
$K^0/\overline{K^0}$ & - & 2.90$\pm$0.24 & 2.88$\pm$0.03 \\
$p/\overline{p}$ & 2.30$\pm$0.20 & 2.50$\pm$0.29 & 3.00$\pm$0.09 \\ \hline
MLLA($\Lambda=200$MeV) & 3.02 & 3.46 & 3.74 \\ \hline
\end{tabular}
\caption{The energy evolution of the peak positions
in $\xi$. Also shown are predictions obtained using the limiting
spectra calculation of MLLA.\label{Peak_position}}
\end{center}
\end{table}

\begin{figure}[t]
{\centerline{\epsfysize=10cm\epsfbox{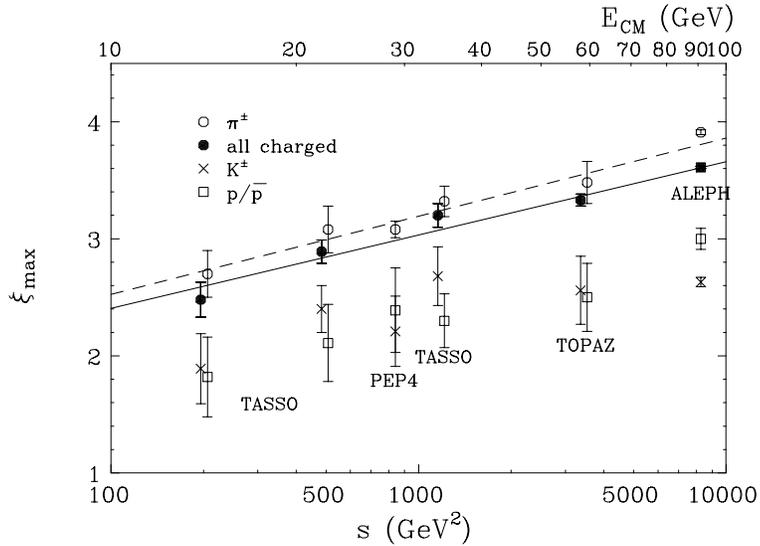}}}
\caption{The peak position in $\xi$ as a function of the square of the
  center-of-mass energy. Solid line shows the linear fit to the peak
  positions of charged particles while dashed line to those of charged
  pions.\label{Energy_Evolution}}
\end{figure}

{}From eq.~\ref{eq:Peak},
the MLLA prediction of peak position can be approximated
in the linear form
$a{\rm ln}E + b$ where $a$ to be 0.5.
If the gluon coherence effect is not taken into account (phase
space only), $a$ should become 1.0. To obtain the value of $a$ from the
measured peak positions, a linear fit
is performed to the peak positions as a function of $\sqrt{s}$
for all charged particles measured by
TASSO\cite{TASSO}, TOPAZ, and ALEPH as shown in
Fig.~\ref{Energy_Evolution}.
{}From the fit, the value of
$a$ is obtained to be $0.54 \pm 0.05$. A similar fit is also done to the peak
positions for charged pions (TASSO, PEP4/TPC, TOPAZ and ALEPH) and $a$ is
measured to be $0.58 \pm 0.04$. These values are consistent with the
MLLA predictions.

\section{Summary}
The inclusive cross sections are measured as a function of
$\xi=ln(1/x_p)$ for all charged particles and for each of $\pi^{\pm}$,
$K^{\pm}$, $p/\overline{p}$ and $K^0/\overline{K^0}$ in the hadronic
events taken at
$\sqrt{s}$=58GeV. The cross sections were
compared with the MLLA calculation assuming LPHD.
The MLLA formula describes the observed distributions very well for
all particle species over the wide momentum range.

By this comparison, $Q_0$ and $\Lambda$ in the MLLA expression are
determined for each particle species.
The determined $Q_0$ for each of the particle species is close to
the mass of the particle, while the $\Lambda$ is almost constant for these
particles. The $Q_0$ values are also determined with
$\Lambda$ fixed at
200MeV. The obtained $Q_0$ also coincides with the mass of
each particle. The same analysis is carried out for
the measurements at $\sqrt{s}=29GeV$ by PEP4/TPC and similar tendency is
observed.
These results show that
the momentum distribution of particles is identical to that
of partons at the end-point of the parton shower evolution independently
of the energy scale.
This supports the MLLA + LPHD conjecture.

The energy evolution of the peak position in $d\sigma/d\xi$ is
studied by comparing our data with the measurements by PEP4/TPC and ALEPH.
The measured peak positions of light particles are well reproduced by the
limiting spectra calculation.
The measured peak positions are fitted to a linear function of
log of the beam energy with the measurements
at other energies. The obtained slope of
the linear function is consistent with the MLLA prediction, while it
excludes models without the gluon coherence effect.

\newpage
\bibstyle{ieee}


\begin{thebibliography}{99}
\bibitem{PTA} PTA collaboration was formed to study the energy
  dependence of QCD. This is the joint effort of PEP4/TPC,
  TOPAZ and ALEPH collaborations.
\end{thebibliography}
\end{document}